\documentclass[
  aps,
  prd,
  nofootinbib,
  nobibnotes,
  superscriptaddress,
  twocolumn
]{revtex4-1}

\pdfoutput=1

\usepackage{amssymb,amsmath}
\usepackage{xcolor}
\usepackage{graphicx}
\usepackage{hyperref}
\usepackage{float} 
\usepackage[T1]{fontenc}
\usepackage{braket}

\newcommand{\nn}{\nonumber}

\begin{document}

\title{Reappraisal of SU(3)-flavor breaking in $B\rightarrow DP$}

\author{Jonathan Davies}
\email{jonathan.davies-7@manchester.ac.uk}
\affiliation{Department of Physics and Astronomy, University of Manchester, Manchester M13 9PL, United Kingdom}

\author{Stefan Schacht}
\email{stefan.schacht@manchester.ac.uk}
\affiliation{Department of Physics and Astronomy, University of Manchester, Manchester M13 9PL, United Kingdom}

\author{Nicola Skidmore}
\email{nicola.skidmore@cern.ch}
\affiliation{Department of Physics, University of Warwick, Coventry CV4 7AL, United Kingdom}

\author{Amarjit Soni}
\email{adlersoni@gmail.com}
\affiliation{Physics Department, Brookhaven National Laboratory, Upton, NY 11973, USA}

\begin{abstract}
In light of recently found deviations of the experimental data from predictions from QCD factorization for $B_{(s)}\rightarrow D_{(s)}P$ decays, where $P=\{\pi,K\}$, we systematically probe the current status of the SU(3)$_F$ expansion from a fit to experimental branching ratio data without any further theory input. We find that the current data are in agreement with the power counting of the SU(3)$_F$ expansion. While the SU(3)$_F$ limit is excluded at $>5\sigma$, amplitude-level SU(3)$_F$-breaking contributions of $\sim 20\%$ suffice for an excellent description of the data. SU(3)$_F$ breaking is needed in tree ($>5\sigma$) and color-suppressed tree ($2.4\sigma$) diagrams.  We are not yet sensitive to SU(3)$_F$ breaking in exchange diagrams. From the underlying SU(3)$_F$ parametrization we predict the unmeasured branching ratios $\mathcal{B}(\overline{B}_s^0\rightarrow \pi^- D^+) = 2 \mathcal{B}(\overline{B}_s^0\rightarrow \pi^0 D^0) = [0.3, 7.2] \times 10^{-6}$ of suppressed decays that can be searched for at the LHCb experiment.
\end{abstract}

\maketitle

\section{Introduction \label{sec:intro}}

Theoretical predictions for the branching ratios of $B_{(s)}\rightarrow D_{(s)}P$, $P=\{\pi,K\}$ and $B_{(s)}\rightarrow D_{(s)}V$, $V=\{\rho, K^*\}$ decays based on QCD factorization~(QCDF) show a significant disagreement with the experimental data \cite{Bordone:2020gao, Cai:2021mlt}. This resulted in a lot of renewed interest in these decays, both theoretically~\cite{Gershon:2021pnc, Piscopo:2023opf, Lenz:2022pgw, Iguro:2020ndk, Endo:2021ifc, Fleischer:2021cct, Beneke:2021jhp, Bordone:2021cca}
and experimentally~\cite{Belle:2022afp, LHCb:2023eig, Dib:2023vot, Belle:2021udv, LHCb:2021qbv}. 
Also in other hadronic $B$ decays anomalies have been seen~\cite{Berthiaume:2023kmp, Bhattacharya:2022akr, Amhis:2022hpm, Biswas:2023pyw, Brod:2014bfa, Jager:2017gal, Lenz:2019lvd}.
In the case of the theoretically clean $\overline{B}_s^0\rightarrow \pi^- D_s^{+}$ and $\overline{B}^0 \rightarrow K^- D^{+}$ decays, which are free from penguin and annihilation topologies and dominated by color-allowed tree processes, this disagreement is sizable; see the comparison of theoretical and experimental results in Table~\ref{tab:comparison}. It is important to note that theoretical predictions based on QCDF for the ratios of the branching fractions, 
\begin{align}
\mathcal{R}_{s/d}^{P(V)}=\frac{\mathcal{B}(\overline{B}_s^0\rightarrow \pi^- D_s^{(*)+})}{\mathcal{B}(\overline{B}^0 \rightarrow K^- D^{(*)+})}\,,  \label{eq:SU3-ratio}
\end{align}
are consistent with the data \cite{Bordone:2020gao} implying a universal effect across $b\rightarrow c\overline{u}q$ weak transitions. 

Decays involving $b\rightarrow c\overline{u}q$ weak transitions are used extensively for measurements of CP violation in the Standard Model (SM). 
Within the SM the Cabibbo Kobayashi Maskawa (CKM) phase $\gamma$ can be measured cleanly via the interference between the amplitudes of $b\rightarrow c\overline{u}s$ and $b\rightarrow u\overline{c}s$ weak transitions with a theoretical uncertainty of $\lesssim 10^{-7}$ \cite{Brod:2013sga}; 
the methodologies to do so include the 
Gronau London Wyler~\cite{Gronau:1990ra, Gronau:1991dp}, 
Atwood Dunietz Soni~\cite{Atwood:1996ci, Atwood:2000ck}, and 
Bondar Poluektov Giri Grossman Soffer Zupan method~\cite{Bondar:2002, Giri:2003ty, Belle:2004bbr, Bondar:2005ki, Ceccucci:2020cim}, see also the overview in Sec.~77 of Ref.~\cite{ParticleDataGroup:2022pth}.
Recent developments regarding unbinned extractions of $\gamma$ can be found in 
Refs.~\cite{Poluektov:2017zxp,Backus:2022xhi, Lane:2023iak}.

Model-independent bounds on non-SM effects in these decays allow a modification to the direct experimental extraction of the CKM phase $\gamma$ by up to $10^{\circ}$~\cite{Lenz:2019lvd} where the current, global, direct experimental determination of $\gamma$ has an uncertainty of approximately~$5^{\circ}$~\cite{CKM21} with LHCb to reach $\sim 0.35^{\circ}$ precision with its final dataset of Upgrade~II~\cite{LHCbUtwo}. 

$B\rightarrow DP$ decays have been studied for a long time.  
The sensitivity to mixing phases and $\gamma$ have been further explored in Refs.~\cite{Dunietz:1997in, Fleischer:2003yb, Fleischer:2003ai}.
Implications of rescattering effects were looked at in Refs.~\cite{Chua:2007qw, Cheng:2004ru, Chua:2001br, Endo:2021ifc}.
QCD factorization and soft-collinear effective theory have been applied in Refs.~\cite{Beneke:2000ry, Bauer:2001cu, Keum:2003js, Huber:2016xod, Cai:2021mlt, Bordone:2020gao, Mantry:2003uz}, and the factorization-assisted topological (FAT) approach in Ref.~\cite{Zhou:2015jba}.
Flavor-symmetry based methods have been used in Refs.~\cite{Zeppenfeld:1980ex, Savage:1989ub, Gronau:1995hm, Chiang:2007bd, Jung:2009pb, Fleischer:2010ca, Kenzie:2016yee, Kitazawa:2018uvd, Colangelo:2005hh, Grinstein:1996us, Neubert:2001sj, Xing:2001nj, Cheng:2001sc, Wolfenstein:2003pc,Chiang:2002tv, Xing:2003fe, Yamamoto:1994pi, Xing:1995sn, Wolfenstein:2003pc, Kim:2004hx}. 
The most recent fits in plain SU(3)$_F$ have been performed in Refs.~\cite{Chiang:2007bd, Colangelo:2005hh,Colangelo:2006kx}.
Reference~\cite{Bordone:2020gao} also looks at SU(3)$_F$ in the context of QCDF.  
New physics sensitivities have recently been explored in Ref.~\cite{Iguro:2020ndk}.

In this paper, in light of the deviations of branching ratio data from QCDF, we analyze systematically the quality of the SU(3)$_F$ expansion in $B_{(s)}\rightarrow D_{(s)}P$ decays, with a methodology similar to the one used in Ref.~\cite{Muller:2015lua} for charm decays. We employ the SU(3)$_F$ expansion derived in Ref.~\cite{Gronau:1995hm}, and include SU(3)$_F$-breaking effects purely through topological diagrams that we extract from the data.
In Sec.~\ref{sec:notation} we introduce our notation, recapitulate the SU(3)$_F$ decomposition and introduce the considered measures of SU(3)$_F$ breaking. We show our numerical results in Sec.~\ref{sec:numerics} and conclude in 
Sec.~\ref{sec:conclusions}

\begin{table*}[t]
\centering
\begin{tabular}{|c|c|c|c|c|c|}
\hline\hline
Observable &  Predictions~I~\cite{Bordone:2020gao} & Predictions II~\cite{Cai:2021mlt} &  Experiment~\cite{ParticleDataGroup:2022pth} &  Deviations~I & Deviations~II \\\hline
$\mathcal{B}(\bar{B}_s^0\rightarrow \pi^- D_s^+)$ & 
 $(4.42\pm 0.21)\cdot 10^{-3}$ & 
  $(4.61^{+0.23}_{-0.39})\cdot 10^{-3}$ &
 $(2.98\pm 0.14)\cdot 10^{-3}$ &
 $ 5.8\sigma$ & $ 4.6\sigma$ \\
 $\mathcal{B}(\bar{B}^0\rightarrow  K^- D^+)$   & 
 $(3.26\pm 0.15)\cdot 10^{-4}$ &  
 $(3.48^{+0.14}_{-0.28})\cdot 10^{-4}$ & 
 $(2.05\pm 0.08)\cdot 10^{-4}$  &  
    $ 7.1\sigma$ & $6.1\sigma$ \\\hline
\end{tabular}
\caption{Comparison of current QCDF predictions and experimental measurements for tree-dominated branching ratios of $B\rightarrow DP$ decays. For the calculation of the deviations, we symmetrize the errors if applicable and divide the quoted theoretical $\bar{B}_s^0$ branching ratios by $1-y_s^2$~\cite{DeBruyn:2012wj}, see Eq.~(\ref{eq:ys}).
\label{tab:comparison}}
\end{table*}

\section{Notation and measures of SU(3)$_F$ breaking \label{sec:notation}}

\subsection{Parametrizations}

We use the normalization 
\begin{align}
\mathcal{B}(B\rightarrow PP') &= \vert \mathcal{A}\vert^2 \cdot \Phi\,,\quad
\Phi = \frac{\tau_B \vert \mathbf{p}\vert}{8\pi m_B^2}\,, \label{eq:phase-space}
\end{align} 
with the phase space factor $\Phi$ and the three-momentum
\begin{align}
\vert \mathbf{p}\vert &= \frac{ \sqrt{
		\left(m_B^2 - (m_P - m_{P'})^2 \right) 
		\left(m_B^2 - (m_P + m_{P'})^2 \right)
		   }}{2 m_B},
\end{align}
where $m_X$ are the particle masses and $\tau_B$ is the lifetime of the $B$ meson. 

For the needed CKM elements we use the Wolfenstein expansion, in $\lambda$, as follows~\cite{Charles:2004jd} 
\begin{align}
\lambda_d \equiv V_{cb} V_{ud}^* &= A\lambda^2(1 - \lambda^2/2)  \,,\\
\lambda_s \equiv V_{cb} V_{us}^* &= A\lambda^3  \,. 
\end{align}

\begin{table}[t]
\centering
\begin{tabular}{|c|c|c|c|c|c|c|c|c|c|}
\hline\hline

$\mathcal{A}(d)$ & $T$ & $C$ & $E$ & $T_1$ & $T_2$ & $C_1$ & $C_2$ & $E_1$ & $E_2$ \\\hline	

\multicolumn{10}{|c|}{$\sim V_{cb} V_{ud}^* = \mathcal{O}(\lambda^2)$}\\\hline

 $B^-\rightarrow \pi^- D^0$ & $-1$ & $-1$ & 0 & 0 & 0 & 0 & 0 & 0 & 0 \\ 

 $\overline{B}^0\rightarrow \pi^- D^+$ & $-1$ & 0 & $-1$ & 0 & 0 & 0 & 0 & 0 & 0 \\ 

 $\overline{B}^0\rightarrow  \pi^0 D^0$  & 0 &  $-\frac{1}{\sqrt{2}}$ & $\frac{1}{\sqrt{2}}$  &  0 & 0 & 0 & 0 & 0 & 0 \\

 $\overline{B}^0\rightarrow K^- D_s^+$ & 0 & 0 & $-1$ & 0 & 0 & 0 & 0 & 0 & $-1$ \\

 $\overline{B}_s^0\rightarrow K^0 D^0$ & 0 & $-1$ & 0 & 0 & 0 & 0 & $-1$  & 0 & 0 \\ 

 $\overline{B}_s^0\rightarrow \pi^- D_s^+$ & $-1$ & 0 & 0 & 0 & $-1$ & 0 & 0 & 0 & 0 \\\hline

\multicolumn{10}{|c|}{$\sim V_{cb} V_{us}^* = \mathcal{O}(\lambda^3)$}\\\hline

 $B^-\rightarrow K^- D^0$ & $-1$ & $-1$ & 0 & $-1$ & 0 & $-1$ & 0 & 0 & 0 \\

 $\overline{B}^0 \rightarrow K^- D^+$  & $-1$ & 0 & 0 & $-1$ & 0 & 0 & 0 & 0 & 0 \\

 $\overline{B}^0\rightarrow \overline{K}^0 D^0$   & 0 & $-1$ & 0 & 0 & 0 & $-1$ & 0 & 0 & 0  \\

 $\overline{B}_s^0\rightarrow \pi^- D^+$  & 0 & 0 & $-1$ & 0 &  0 & 0 & 0 & $-1$ & 0  \\

 $\overline{B}_s^0\rightarrow \pi^0 D^0$  & 0 & 0 & $\frac{1}{\sqrt{2}}$ & 0 & 0 & 0 & 0 & $\frac{1}{\sqrt{2}}$ & 0 \\

 $\overline{B}_s^0 \rightarrow K^- D_s^+$  & $-1$ & 0 & $-1$ & $-1$ & $-1$ & 0 & 0 & $-1$ & $-1$ \\\hline\hline 

\end{tabular}
\caption{SU(3)$_F$ decomposition. The table is adapted from Ref.~\cite{Gronau:1995hm}.
\label{tab:SU3-decomposition}}
\end{table}

We use the topological SU(3)$_F$ decomposition given in Ref.~\cite{Gronau:1995hm} which we summarize for convenience in Table~\ref{tab:SU3-decomposition}.
For our fit of the parametrization to the data, we use the following combination of parameters:
\begin{align}
&T\,,\,
\left\vert \frac{C}{T}\right\vert\,,\,
\left\vert \frac{E}{T}\right\vert\,,\,
\phi_C\,,\,
\phi_E\,, 
\left\vert \frac{T_1}{T}\right\vert\,,\,
\left\vert \frac{T_2}{T}\right\vert\,,\,
\phi_{T_1}\,,\,
\phi_{T_2}\,,\nn\\
&\left\vert \frac{C_1}{C}\right\vert\,,\,
\left\vert \frac{C_2}{C}\right\vert\,,\,
\phi_{C_1}\,,\,
\phi_{C_2}\,, 
\left\vert \frac{E_1}{E}\right\vert\,,\,
\left\vert \frac{E_2}{E}\right\vert\,,\,
\phi_{E_1}\,,\,
\phi_{E_2}\,, \label{eq:su3-parameters}
\end{align}
with
\begin{align}
&\phi_C \equiv \mathrm{arg}(C/T)\,,\,
\phi_E \equiv \mathrm{arg}(E/T)\,,\,
\phi_{T_i} \equiv \mathrm{arg}(T_i/T)\,,\,\nn\\
&\phi_{C_i} \equiv \mathrm{arg}(C_i/T)\,,\,
\phi_{E_i} \equiv \mathrm{arg}(E_i/T)\,.
\end{align}
Furthermore, we choose $T$ to be real and positive without loss of generality. 
All topological diagrams carry strong phases only.

SU(3)$_F$ is an approximate symmetry of the QCD Lagrangian that becomes exact in the limit $m_u=m_d=m_s$. Effects from corrections to this limit that account for $m_s\neq m_d$ can be incorporated in a systematic fashion with perturbation theory in terms of mass insertions of the form $\sim (m_s-m_d)\bar{s}s$~\cite{Gronau:1995hm, Muller:2015lua, Jung:2009pb}. The diagrams $T_i$, $C_i$, and $E_i$ have the same flavor and color flow as the corresponding diagrams $T$, $C$, and $E$, except that they each include one mass insertion on one of the strange quark fermion lines (see Fig.~2 in Ref.~\cite{Gronau:1995hm}). In this study we include effects at first order in the expansion in SU(3)$_F$ breaking only. Second order corrections would come from diagrams that include two strange quark mass insertions and are beyond the scope of this study.

We also fit the isospin decompositions of $B\rightarrow \pi D$ and $B\rightarrow KD$ to the available data.
These are given as~\cite{Gronau:1995hm}
\begin{align}
A_{\pi D}^{-0} \equiv A(B^-\rightarrow  \pi^- D^0)      &= A_{3/2}\,, \\
A_{\pi D}^{-+} \equiv A(\bar{B}^0\rightarrow \pi^- D^+) &= \frac{2}{3} A_{1/2} + \frac{1}{3} A_{3/2}\,,\\
A_{\pi D}^{00} \equiv A(\bar{B}^0\rightarrow \pi^0 D^0) &= -\frac{\sqrt{2}}{3} A_{1/2} + \frac{\sqrt{2}}{3} A_{3/2}\,,\\
A_{K D}^{-0} \equiv A(B^-\rightarrow K^- D^0 ) &= A_1\,, \\
A_{K D}^{-+} \equiv A(\bar{B}^0\rightarrow K^- D^+ ) &= \frac{1}{2} A_1 + \frac{1}{2} A_0\,, \\
A_{K D}^{00} \equiv A(\bar{B}^0\rightarrow  \bar{K}^0 D^0) &= \frac{1}{2} A_1 - \frac{1}{2} A_0\,,
\end{align}
with the parameters
\begin{align}
\vert A_{1/2}\vert\,,
\vert A_{3/2}\vert\,,
\cos(\mathrm{arg}(A_{1/2}/ A_{3/2})) 
\end{align}
and
\begin{align}
\vert A_{0}\vert\,,
\vert A_{1}\vert\,,
\cos(\mathrm{arg}(A_0/ A_1))\,, 
\end{align}
respectively.
For the isospin systems there are simple known closed-form expressions for the parameters \cite{Chiang:2002tv} 
\begin{align}
&\left|\frac{A_{1/2}}{A_{3/2}} \right| = \sqrt{\frac{3\left|A_{\pi D}^{-+}\right|^2+3\left|A_{\pi D}^{00}\right|^2-\left|A_{\pi D}^{-0}\right|^2}{2\left|A_{\pi D}^{-0}\right|^2}}\,,\\
&\cos\left(\mathrm{arg}\left(\frac{A_{1/2}}{A_{3/2}}\right)\right) = \nn\\ & \quad \frac{3\left|A_{\pi D}^{-+}\right|^2+\left|A_{\pi D}^{-0}\right|^2-6\left|A_{\pi D}^{00}\right|^2}{2\sqrt{2}\left|A_{\pi D}^{-0}\right|\sqrt{3\left|A_{\pi D}^{-+}\right|^2+3\left|A_{\pi D}^{00}\right|^2-\left|A_{\pi D}^{-0}\right|^2}}\,,
\end{align}
and \cite{Xing:2003fe}
\begin{align}
&\left|\frac{A_0}{A_1}\right| =  \sqrt{\frac{2\left|A_{KD}^{-+}\right|^2+2\left|A_{KD}^{00}\right|^2-\left|A_{KD}^{-0}\right|^2}{\left|A_{KD}^{-0}\right|^2}}\,, \\
&\cos\left(\mathrm{arg}\left(\frac{A_0}{A_1}\right)\right) =  \nn\\& \quad \frac{\left|A_{KD}^{+-}\right|^2-\left|A_{K D}^{00}\right|^2}{\left|A_{KD}^{- 0}\right|\sqrt{2\left|A_{KD}^{-+}\right|^2+2\left|A_{KD}^{00}\right|^2-\left|A_{KD}^{-0}\right|^2}}\,,
\end{align}
respectively.

\subsection{Measures of SU(3)$_F$ breaking \label{sec:measures-su3}}

As parametrization-dependent measures of SU(3)$_F$ breaking we use the parameters of Eq.~(\ref{eq:su3-parameters})
\begin{align}
&\left\vert \frac{T_1}{T}\right\vert\,,\,
\left\vert \frac{T_2}{T}\right\vert\,,\,
\left\vert \frac{C_1}{C}\right\vert\,,\,
\left\vert \frac{C_2}{C}\right\vert\,,\,
\left\vert \frac{E_1}{E}\right\vert\,,\,
\left\vert \frac{E_2}{E}\right\vert\,.
\end{align}
As an alternative, parametrization-independent measure of SU(3)$_F$ breaking we also consider the 
known SU(3)$_F$ sum rules~\cite{Gronau:1995hm} between two decay channels and calculate the splitting 
around their average value
\begin{align}
\varepsilon(A_1/A_2) &\equiv \left|\frac{ \vert A_1\vert - \frac{1}{2}(\vert A_1\vert + \vert A_2\vert) }{\frac{1}{2}(\vert A_1\vert + \vert A_2\vert)}\right|
= \left|\frac{\vert A_1\vert - \vert A_2\vert}{\vert A_1\vert + \vert A_2\vert} \right| \\
&= \left|\frac{1 - \vert A_1/A_2\vert}{1 + \vert A_1/A_2\vert}\right|\,,
\end{align}
in analogy to the splitting of spectral lines around a symmetry limit. 
Note that $\varepsilon(A_1/A_2)$ is symmetric with respect to the interchange of $A_1$ and $ A_2$. 

To that end, we define the ratios
\begin{align}
s_1 &\equiv \frac{1}{\sqrt{2}} \frac{
        \vert \mathcal{A}(\bar{B}_s^0\rightarrow \pi^- D^+)\vert}{ 
	\vert \mathcal{A}(\bar{B}_s^0\rightarrow \pi^0 D^0)\vert}\,, \label{eq:sum-rule-1}\\
s_2 &\equiv  \frac{\vert \mathcal{A}(\bar{B}^0\rightarrow K^- D_s^+)\vert/\vert\lambda_d\vert}{ \vert \mathcal{A}(\bar{B}_s^0\rightarrow \pi^- D^+ )\vert/\vert\lambda_s\vert }\,,  \label{eq:sum-rule-2}\\
s_3 &\equiv \frac{\vert \mathcal{A}( B^-\rightarrow \pi^- D^0 )\vert/\vert\lambda_d\vert }{\vert \mathcal{A}(B^-\rightarrow K^- D^0 )\vert/\vert\lambda_s\vert} \,,\label{eq:sum-rule-3}\\
s_4 &\equiv  \frac{\vert \mathcal{A}( \bar{B}^0\rightarrow \pi^- D^+ )\vert/\vert\lambda_d\vert}{\vert \mathcal{A}(\bar{B}_s^0\rightarrow K^- D_s^+)\vert/\vert\lambda_s\vert} \,, \label{eq:sum-rule-4}  \\
s_5 &\equiv \frac{ \vert \mathcal{A}(\bar{B}_s^0\rightarrow K^0 D^0)\vert/\vert\lambda_d\vert}{\vert \mathcal{A}(\bar{B}^0\rightarrow \bar{K}^0 D^0)\vert/\vert\lambda_s\vert}\,, \label{eq:sum-rule-5} \\
s_6 &\equiv \frac{\vert \mathcal{A}( \bar{B}_s^0\rightarrow \pi^- D_s^+)\vert/\vert\lambda_d\vert }{\vert \mathcal{A}(\bar{B}^0\rightarrow K^-D^+ )\vert/\vert\lambda_s\vert}\,, \label{eq:sum-rule-6}
\end{align}
where in the SU(3)$_F$ limit (including isospin) 
\begin{align}
s_i = 1\,, \label{eq:sum-rules}
\end{align}
such that in this limit
\begin{align}
\varepsilon_i \equiv \varepsilon(s_i) = 0\,.
\end{align}
Note that additional amplitude sum rules that connect more than two decay channels do exist~\cite{Gronau:1995hm}.

\section{Numerical Results \label{sec:numerics}}

\begin{table}[t]
\begin{center}
\begin{tabular}{|c|c|c|}
\hline\hline

Observable & Value & Ref. \\\hline

\multicolumn{2}{|c|}{$\sim V_{cb} V_{ud}^* = \mathcal{O}(\lambda^2)$} & \\\hline

$\mathcal{B}(B^-\rightarrow \pi^- D^0)$ & $(4.61\pm 0.10)\cdot 10^{-3} $ & \cite{ParticleDataGroup:2022pth} \\ 

$\mathcal{B}(\overline{B}^0\rightarrow \pi^- D^+)$  & $(2.51\pm 0.08)\cdot 10^{-3}$  & \cite{ParticleDataGroup:2022pth}\\ 

$\mathcal{B}(\overline{B}^0\rightarrow  \pi^0 D^0)$  &  $(2.67\pm 0.09)\cdot 10^{-4}$ &  \cite{ParticleDataGroup:2022pth}\\

$\mathcal{B}(\overline{B}^0\rightarrow K^- D_s^+)$ &  $(2.7\pm 0.5)\cdot 10^{-5}$ & \cite{ParticleDataGroup:2022pth} \\

$\mathcal{B}(\overline{B}_s^0\rightarrow K^0 D^0)$ &  $(4.3\pm 0.9)\cdot 10^{-4}$  & \cite{ParticleDataGroup:2022pth}\\ 

$\mathcal{B}(\overline{B}_s^0\rightarrow \pi^- D_s^+)$ & $(2.98\pm 0.14)\cdot 10^{-3}$ &  \cite{ParticleDataGroup:2022pth}\\\hline

\multicolumn{2}{|c|}{$\sim V_{cb} V_{us}^* = \mathcal{O}(\lambda^3)$} & \\\hline

$\mathcal{B}(B^-\rightarrow K^- D^0)$ & $(3.64\pm 0.15)\cdot 10^{-4}$ & \cite{ParticleDataGroup:2022pth}\\

$\mathcal{B}(\overline{B}^0 \rightarrow K^- D^+)$  & $(2.05 \pm 0.08)\cdot 10^{-4}$ & \cite{ParticleDataGroup:2022pth} \\

$ \mathcal{B}(\overline{B}^0\rightarrow \overline{K}^0 D^0)$ &  $(5.2\pm 0.7 )\cdot 10^{-5}$  & \cite{ParticleDataGroup:2022pth}\\

$ \mathcal{B}(\overline{B}_s^0\rightarrow \pi^- D^+)$  & n.a. &  \cite{ParticleDataGroup:2022pth} \\

$\mathcal{B}(\overline{B}_s^0\rightarrow \pi^0 D^0)$  & n.a. & \cite{ParticleDataGroup:2022pth} \\

$\mathcal{B}(\overline{B}_s^0 \rightarrow K^- D_s^+)$  &  $(1.94\pm 0.21)\cdot 10^{-4}$   
 & \cite{Fleischer:2021cct}  \\\hline\hline 
\end{tabular}
\end{center}
\caption{Experimental input data, see text for details.
 \label{tab:experimental}}
\end{table}

\begin{table}[t]
\centering
\begin{tabular}{|c|c|}
\hline\hline
Parameter & Value  \\\hline
$T$ &  $[1.36, 1.63] \cdot 10^{-5}$~GeV\\
$\vert C/T\vert$ & [0.39, 0.63]\\
$\vert E/T\vert$ & [0.07, 0.16] \\
$\phi_C$ &  $[-84, -64]^{\circ}$,  $[64, 84]^{\circ}$ \\
$\phi_E$ &  $[-165, -79]^{\circ}$, $[75, 165]^{\circ}$ \\
\hline \hline
$\vert T_1/T\vert$ & [0.10, 0.30] \\
$\vert T_2/T\vert$ &  \textcolor{gray}{[0.0, 0.3]} \\ 
$\phi_{T_1}$ &  $[-85, 85]^{\circ}$\\
$\phi_{T_2}$ &  $\textcolor{gray}{[-180.0, 180.0]^{\circ}}$ \\
$\vert C_1/C\vert$ &  \textcolor{gray}{[0.00, 0.30]}\\
$\vert C_2/C\vert$ &   [0.02, 0.30] \\
$\phi_{C_1}$ & $\textcolor{gray}{[-180.0, 180.0]^{\circ}}$ \\
$\phi_{C_2}$ &  $[-180, -22]^{\circ}$, $[22, 180]^{\circ}$\\
$\vert E_1/E\vert$ &  \textcolor{gray}{[0.0, 0.3]}  \\
$\vert E_2/E\vert$ &   \textcolor{gray}{[0.0, 0.3]} \\
$\phi_{E_1}$ & $\textcolor{gray}{[-180.0, 180.0]^{\circ}}$ \\
$\phi_{E_2}$ &  $\textcolor{gray}{[-180.0, 180.0]^{\circ}}$ \\\hline\hline
\end{tabular}
\caption{Numerical results for the topological matrix elements. SU(3)$_F$ breaking is constrained not to exceed 30$\%$, see Eqs.~(\ref{eq:SU3-breaking-1})--(\ref{eq:SU3-breaking-4}).
The $\Delta\chi^2$ profiles are non-Gaussian and flat around the minimum; therefore, we only show $1\sigma$ ranges without central point.
In light gray we show parameters to which we do not have sensitivity and therefore just reproduce our corresponding theoretical assumption. \label{tab:numerical-results}}
\end{table}

\begin{table}[t]
\centering
\begin{tabular}{|c|c|c|c|}
\hline\hline
Hypothesis &  $\Delta \chi^2$ & $\Delta$dof & Significance of rejection  \\\hline
   $T_i=C_i=E_i=0$  & $103.2$  & $12$  & $>5\sigma$ \\
   $T_i=0$ & $37.3$ & $4$ & $>5\sigma$ \\
   $C_i=0$ & $12.0$ & $4$ & $2.4\sigma$\\
   $E_i=0$ & $0.0$ & $4$ & $0\sigma$\\
   $T_i=C_i=0$ & $91.4$ & 8 & $>5\sigma$\\
   $T_i=E_i=0$ & $50.8$ & 8 & $>5\sigma$\\
   $C_i=E_i=0$ & $12.5$ & 8 & $1.5\sigma$\\\hline\hline
\end{tabular}
\caption{Significance of rejection of benchmark hypotheses compared to the null hypothesis of a complete fit with $\delta_X=0.3$. $\Delta$dof indicates the difference of the number of degrees of freedom for the considered nested hypothesis compared to the null hypothesis. \label{tab:hypothesis-tests}
}
\end{table}

\begin{table}[t]
\begin{tabular}{|c|c|}
\hline\hline

$\varepsilon_3$ & $0.10 \pm 0.01$\\

$\varepsilon_4$ & $0.10 \pm 0.03$   \\

$\varepsilon_5$ & $0.20 \pm  0.06$
\\
$\varepsilon_6$ & $0.06\pm 0.02$ 
\\\hline\hline

\end{tabular}
\caption{Numerical results for the parametrization-independent measures of SU(3)$_F$ breaking, resulting from $\Delta\chi^2$ scans.
We do not show results for $\varepsilon_{1,2}$ as these contain yet unmeasured branching ratios, see Eqs.~(\ref{eq:sum-rule-1}) and (\ref{eq:sum-rule-2}).
\label{tab:sum-rule-tests}}
\end{table}

\begin{table}[t]
\centering
\begin{tabular}{|c|c|}
\hline\hline
Parameter & Value  \\\hline
$\vert A_{1/2}/ A_{3/2}\vert$ & $0.69\pm0.03$\\
$\left|\mathrm{arg}(A_{1/2}/ A_{3/2})\right|$ &  $(30^{+1}_{-2})^{\circ}$ \\
$\vert A_0/ A_1\vert$ & $0.72\pm0.06$\\
$\left|\mathrm{arg}(A_0/ A_1)\right|$ & $(51 \pm 4)^{\circ}$ \\\hline\hline
\end{tabular}
\caption{Numerical results for isospin matrix elements.
\label{tab:isospin-results}
}
\end{table}

\begin{figure}[t]
    \centering
    \includegraphics[width=0.49\textwidth]{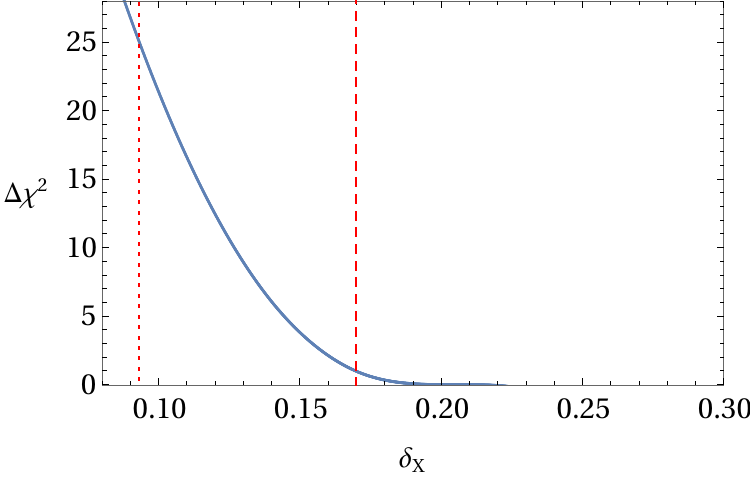}
    \caption{$\Delta\chi^2$ for varying degrees of allowed SU(3)$_F$ breaking $\delta_X$. $\Delta\chi^2=1\, (25)$,~\emph{i.e.},~$1\, (5)\sigma$, corresponds to $\delta_X=0.17(0.09)$,
    indicated by the dashed(dotted) line, and $\Delta\chi^2<1\times 10^{-3}$ is reached for $\delta_X>0.2$.
    \label{fig:chi2-deltaX}}
\end{figure}

The current experimental status of the relevant branching ratio measurements is shown in Table~\ref{tab:experimental}. For $B_s$ decays, due to $\Delta\Gamma_s\neq 0$, we have to take into account a correction factor for the relation between the \lq\lq{}theoretical\rq\rq{} branching ratio at $t=0$ and the \lq\lq{}experimental\rq\rq{} time-integrated branching ratio~\cite{DeBruyn:2012wj}.
The implications of $\Delta \Gamma_s\neq 0$ for $B_s\rightarrow D_s^{\pm}K^{\mp}$ have been discussed in detail in  Ref.~\cite{Fleischer:2021cct} and we use the 
\lq\lq{}theoretical\rq\rq{} $\mathcal{B}(\overline{B}_s^0 \rightarrow K^- D_s^+)$ as extracted from the experimental data therein. This amounts to a correction of 
$\sim 14\%$ vs.~the \lq\lq{}experimental\rq\rq{} value $\mathcal{B}(\overline{B}_s^0 \rightarrow K^{\mp} D_s^{\pm}) = (2.25\pm 0.12)\cdot 10^{-4}$~\cite{ParticleDataGroup:2022pth}; see Refs.~\cite{Fleischer:2021cct, DeBruyn:2012jp} for details.

The decay $\bar{B}_s^0\rightarrow \pi^- D_s^+$ is flavor specific (see Ref.~\cite{Gershon:2021pnc} for a detailed discussion). Therefore, in this case the correction factor accounting for the nonvanishing $B_s$ width difference 
$\Delta\Gamma_s$ reads $1-y_s^2$~\cite{DeBruyn:2012wj}, with~\cite{HFLAV:2022esi}
\begin{align}
y_s=\Delta\Gamma_s/(2 \Gamma_s) = 0.064\pm 0.0035\,, \label{eq:ys}
\end{align}
and the average $B_s$ width $\Gamma_s$. As $y_s^2\sim 0.004$, and the relative uncertainty of $\bar{B}_s^0\rightarrow \pi^- D_s^+$ is an order of magnitude larger, we neglect the effect of the nonvanishing width difference
for this decay.

The decay channel $\overline{B}_s^0\rightarrow K^0 D^0$ interferes with $B_s^0\rightarrow \overline{K}^0 D^0$ through $B_s$ and kaon mixing. However, the latter decay channel is relatively suppressed by $\mathcal{O}(\lambda^2)$~\cite{Gronau:1995hm}. Therefore, compared to the current relative error of $\sim 20\%$ for $\mathcal{B}(\overline{B}_s^0\rightarrow K^0 D^0)$, we neglect the effect of $\Delta \Gamma_s\neq 0$ in this case, too.

For our global fit, we assume that 
\begin{align}
\vert T_i/T\vert &\leq \delta_X\,, \label{eq:SU3-breaking-1}\\
\vert C_i/C\vert &\leq \delta_X\,, \label{eq:SU3-breaking-2}\\
\vert E_i/E\vert &\leq \delta_X\,, \label{eq:SU3-breaking-3}
\end{align}
where $\delta_X$ is a measure of the allowed SU(3)$_F$ breaking which we set to
\begin{align}
\delta_X &= 0.3\,. \label{eq:SU3-breaking-4}
\end{align}
We have altogether ten experimental data points (two branching ratios not yet being measured) and 17 real theory parameters, of which nine are magnitudes and eight are strong phases. However, we also have six constraints Eqs.~(\ref{eq:SU3-breaking-1})--(\ref{eq:SU3-breaking-3}) on these parameters, and it is indeed nontrivial if the data are in agreement with these constraints.

At the global minimum, for our null hypothesis we find $\chi^2=0$, \emph{i.e.},~a perfect fit to the data, meaning that the data can be explained with SU(3)$_F$ breaking of $\delta_X=30\%$, as we  expect.
To demonstrate the nontriviality of this result, we show in Fig.~\ref{fig:chi2-deltaX}
the resulting $\Delta \chi^2$ profile as a function of $\delta_X$, compared to our null hypothesis.

Our fit results for $\delta_X=0.3$ are presented in 
Tables~\ref{tab:numerical-results}, \ref{tab:hypothesis-tests}, and \ref{tab:sum-rule-tests}.
 All errors are obtained by profiled-$\chi^2$ scans.
Since the SU(3)$_F$ system is underconstrained, with more parameters than observables, it has a high degree of degeneracy. The parameter scans show non-Gaussianity, and appear flat around the minimum while sharply increasing near parameter boundaries. For that reason we quote parameter ranges instead of a central point
with errors.

We note especially that the SU(3)$_F$ limit, which corresponds to $\delta_X=0$ in Fig~\ref{fig:chi2-deltaX} and therefore to a fit with only five parameters,
\begin{align}
&T\,,\,
\left\vert \frac{C}{T}\right\vert\,,\,
\left\vert \frac{E}{T}\right\vert\,,\,
\phi_C\,,\,
\phi_E\,, 
\end{align}
results in a very high $\chi^2$, and is excluded. 
We understand this result in view of the SU(3)$_F$ limit sum rules $\varepsilon_i=0$, see Sec.~\ref{sec:measures-su3}. These sum rules effectively overconstrain the fit and are clearly broken by the data (see Table~\ref{tab:sum-rule-tests}). Therefore, the SU(3)$_F$ limit is inconsistent with the data, and SU(3)$_F$ breaking contributions have to be taken into account. Only then, does the fit have the flexibility to reach $\varepsilon_i\neq 0$ as 
required by the data. However, partial degeneracies arise concerning exactly how the required splitting of the branching ratios encoded in the $\varepsilon_i$ is reached. Considering, for example, the quantity $\varepsilon_3$ which measures the splitting between $\mathcal{B}(B^-\rightarrow \pi^- D^0)$ and $\mathcal{B}(B^-\rightarrow K^-D^0)$, $\varepsilon_3\neq 0$ can be achieved by both $T_1\neq 0$ or $C_1\neq 0$. Analogous observations hold for the other branching ratio splittings. A global fit is necessary in order to resolve these ambiguities.

For completeness, in Table~\ref{tab:isospin-results} we also show updated fits of the $B\rightarrow D\pi$ and $B\rightarrow DK$ isospin systems, which are in agreement with and more precise than previous results in Refs.~\cite{Neubert:2001sj, Xing:2001nj, Cheng:2001sc, Chiang:2002tv, Mantry:2003uz} and \cite{Colangelo:2005hh, Xing:2003fe, Mantry:2003uz}, respectively, thanks to the steady experimental progress~\cite{Belle:2001ccu, CLEO:2002iij, Belle:2002jgk, BaBar:2003rla, BaBar:2006rof, BaBar:2006zod, BaBar:2006ftr, Belle:2008msg, BaBar:2011qjw, LHCb:2013vfg, Belle:2017psv, LHCb:2020hdx, Belle:2021nyg, Belle:2021udv}.

We make the following observations:
\begin{itemize}
\item As already mentioned, the SU(3)$_F$ limit is excluded at $>5\sigma$, mainly triggered by the need for SU(3)$_F$ breaking in the tree diagrams, which is also $>5 \sigma$. SU(3)$_F$ breaking in the $C$ diagrams is needed at $2.4\sigma$, while we are not yet sensitive to SU(3)$_F$ breaking in $E$ diagrams.
\item To further explore and illustrate the different importance of SU(3)$_F$ breaking in the various topologies, we consider scenarios in which SU(3)$_F$ breaking effects in only one topology at a time are taken into account, see the last three lines of Table~\ref{tab:hypothesis-tests}. This demonstrates that SU(3)$_F$ breaking in only $C$ or $E$ does not suffice in order to obtain a good description of the data. The crucial component of the SU(3)$_F$-breaking corrections are the ones in the tree amplitude, because only with $T_i\neq 0$ is the $\chi^2$ reduced to an acceptable level.  
\item Compared to Ref.~\cite{Colangelo:2005hh}, where the topological amplitudes are extracted in the SU(3)$_F$ limit, and to Ref.~\cite{Chiang:2007bd}, where partial SU(3)$_F$ breaking effects are taken into account, we find that the general features of the results for $\vert C/T\vert$ and $\vert E/T\vert$ obtained in \cite{Colangelo:2005hh, Chiang:2007bd} are confirmed when including the first order corrections in full generality; however, the additional parameters result in larger errors for $\vert C/T\vert$.
\item The data can be described perfectly well with SU(3)$_F$ breaking of $\sim20\%$. 
This conclusion holds independently of employing parameter-dependent or parameter-independent measures of SU(3)$_F$ breaking, see Fig.~\ref{fig:chi2-deltaX} and Table~\ref{tab:sum-rule-tests}.
Our observation agrees with the conclusions of studies of SU(3)$_F$ breaking based on factorization~\cite{Bordone:2020gao, Fleischer:2010ca}, where the factorizable SU(3)$_F$ breaking from the decay constants can be estimated as $f_K/f_{\pi}-1\sim 20\%$~\cite{FlavourLatticeAveragingGroupFLAG:2021npn}, see also Ref.~\cite{Jung:2009pb}.
\item The SU(3)$_F$-breaking effects in $T$, $C$, and $E$ do not add up constructively in order
to produce larger SU(3)$_F$-breaking effects. This can be seen from the SU(3)$_F$-breaking measure $\varepsilon_4$, which relates the $U$-spin pair $\overline{B}^0 \to \pi^- D^+$ and $\overline{B_s}^0 \to K^- D_s^+$, see Table~\ref{tab:sum-rule-tests}. 
\item The $B\rightarrow D\pi$ isospin amplitude ratio is still in agreement with the heavy-quark limit estimate~\cite{Neubert:2001sj}
\begin{align}
\vert A_{1/2}/A_{3/2}\vert &\overset{\text{HQ-limit}}{=} 1 + \mathcal{O}(\Lambda_{\mathrm{QCD}}/m_c)\,.
\end{align}
\item The data imply that the ratio $\vert E/T\vert$ is more strongly suppressed than $\vert C/T\vert$. 
However, $\vert C/T\vert$ is rather large compared to the expectation from $1/N_c$ counting~\cite{Gronau:1995hm}. 
The obtained knowledge of $\vert E/T\vert$ is mainly driven by the constraint from  
\begin{align}
&\left| \frac{
    \mathcal{A}(\overline{B}^0\rightarrow K^- D_s^+)
        }{
    \mathcal{A}(\overline{B}^0\rightarrow \pi^- D^+)
    }\right| = \left|\frac{E+E_2}{T+E}\right|\,,
\end{align}
together with $\left|E_2/E\right| < 0.3$. 
\item The phases $\phi_C$ and $\phi_E$ are sizable, indicating large rescattering effects.
The phases of the SU(3)$_F$-breaking topologies are not yet very well known. 
\item Because of lack of data we cannot yet test the isospin relation $s_1=1$; see Eqs.~(\ref{eq:sum-rule-1}, \ref{eq:sum-rules}). From our fit we obtain the corresponding yet unmeasured branching ratios as
\begin{align}
&\quad \mathcal{B}(\bar{B}_s^0(t=0)\rightarrow \pi^- D^+) \nn\\
&= 2 \mathcal{B}(\bar{B}_s^0(t=0)\rightarrow \pi^0 D^0) \nn\\ 
 &= [0.4, 6.0]\cdot 10^{-6}\,. 
\end{align}
The interfering $B_s$ decays into the same final states are in this case of the same order in Wolfenstein-$\lambda$, as it is also the case for $\overline{B}_s^0 \rightarrow K^{\mp} D_s^{\pm}$ decays, see above. 
Estimating the impact from $\Delta \Gamma_s\neq 0$ as $\sim 20\%$, as suggested by the size of the effect for $\mathcal{B}(\overline{B}_s^0 \rightarrow K^- D_s^+)$, we predict
\begin{align}
\mathcal{B}(\bar{B}_s^0\rightarrow \pi^- D^+) &= 2 \mathcal{B}(\bar{B}_s^0\rightarrow \pi^0 D^0)\nn\\
&= [0.3, 7.2]\cdot 10^{-6}\,. \label{eq:br-prediction}
\end{align}
Future measurements of 
$\mathcal{B}(\overline{B}^0_s \to \pi^- D^+)$ and $\mathcal{B}(\overline{B}^0_s \to \pi^0 D^0)$ would allow one to test their isospin symmetry relation and at the same time improve the bounds on $\vert E/T\vert$.
\end{itemize}

\section{Conclusions \label{sec:conclusions}}

Motivated by recently found discrepancies between experimental data and QCD factorization, we study the anatomy of a plain SU(3)$_F$ expansion in $B\rightarrow DP$ decays as extracted from the data, with no further theory assumptions. We find that, although the SU(3)$_F$ limit 
is excluded by $>5\sigma$, the data can be explained with amplitude-level SU(3)$_F$ breaking of 
$\sim 20\%$;~\emph{i.e.},~the SU(3)$_F$ expansion works to the expected level. 

As is well known, SU(3)$_F$ is an approximate symmetry of QCD as the light ($u$, $d$, $s$) quark masses are not equal to each other though they (especially $u$ and $d$) are small compared to $\Lambda_{\mathrm{QCD}}$.
So when predictions made by assuming SU(3)$_F$ are not upheld experimentally,   it often becomes important to estimate the size of SU(3)$_F$ breaking before invoking new physics. 

As one more piece in the puzzle, our results inform future theoretical and experimental work.
By considering the complete system of decays related by the SU(3)$_F$ symmetry, we show 
unambiguously that the underlying issue is not connected to SU(3)$_F$ breaking, as suggested by the ratio in Eq.~(\ref{eq:SU3-ratio}), which is in agreement with the data. 
In a scenario beyond the SM (BSM), any new physics model designed to explain the deviations has therefore to respect the SU(3)$_F$ structure of the SM operators, implying bounds on the hierarchy between BSM couplings to the first and second generations of quarks.

It would be very interesting to complete the picture in the future by testing the isospin sum rule 
between $\mathcal{B}(\bar{B}_s^0\rightarrow \pi^- D^+)$ and $\mathcal{B}(\bar{B}_s^0\rightarrow \pi^0 D^0)$, 
as well as the SU(3)$_F$ sum rule between 
$\mathcal{B}(\bar{B}^0\rightarrow K^- D_s^+)$ and $\mathcal{B}(\bar{B}_s^0\rightarrow \pi^- D^+ )$,
see Eqs.~(\ref{eq:sum-rule-1}, \ref{eq:sum-rule-2}, \ref{eq:sum-rules}). 
To achieve this, measurements of the suppressed decay 
modes $\bar{B}_s^0\rightarrow \pi^-D^+$ and 
$\bar{B}_s^0\rightarrow \pi^0D^0$ are necessary, whose branching ratios we predict in Eq.~(\ref{eq:br-prediction}).

The LHCb Upgrade II experiment has strong prospects for these modes.  Run~5 of the LHC, scheduled for 2035, will see LHCb operating at instantaneous luminosities an order of magnitude greater than previously and accumulating a data sample corresponding to a minimum of 300~fb$^{-1}$. This integrated luminosity, accompanied by improvements to the electromagnetic calorimeter granularity and energy resolution, will provide unprecedented sensitivity to modes with neutral particles in the final state~\cite{LHCbUtwo}.

From our parameter extraction in Table~\ref{tab:numerical-results} it is evident that many of the SU(3)$_F$ breaking parameters are basically unconstrained. Future more precise branching ratio data would allow one to test the pattern of the SU(3)$_F$ anatomy with much more accuracy.
Besides the unmeasured decay channels $\bar{B}_s^0\rightarrow \pi^-D^+$ and $\bar{B}_s^0\rightarrow \pi^0 D^0$, there is a lot of room for improvement left in the channels
$\bar{B}^0\rightarrow K^- D_s^+$, $\bar{B}_s^0\rightarrow K^0D^0$, $\bar{B}^0\rightarrow \bar{K}^0 D^0$ and $\bar{B}_s^0\rightarrow K^-D_s^+$, all of which still have relative branching ratio uncertainties of $>10\%$.
Being a general characteristic of symmetry-based methods, we need improvements in several decay channels in order to obtain a complete picture of the underlying dynamics of $B\rightarrow DP$ decays.

\begin{acknowledgments}
J.D. is supported by the European Research Council under the starting grant  Beauty2Charm 852642.
The work of A.S.~is supported in part by the US DOE Contract No. DE-SC 0012704.
S.S. is supported by a Stephen Hawking Fellowship from UKRI under reference EP/T01623X/1 and the STFC research grants ST/T001038/1 and ST/X00077X/1.
\end{acknowledgments}
\bibliography{draft.bib}
\bibliographystyle{apsrev4-1}

\end{document}